\documentclass[twocolumn,showpacs,preprintnumbers,amsmath,amssymb]{revtex4}
\usepackage{graphicx}

\newcommand{\half}{\mbox{\small $1 \over 2$}}

\newcommand{\bv}{{\bf v}}
\newcommand{\bx}{{\bf x}}
\newcommand{\br}{{\bf r}}

\begin{document}
\title{Linear response formula for open systems}
\author{Onuttom Narayan}
\affiliation{Department of Physics, University of California, Santa Cruz, CA 95064}
\date{\today}
\begin{abstract}
An exact expression for the finite frequency response of open
classical systems coupled to reservoirs is obtained. The result is
valid for any conserved current. No assumption is made about the
reservoirs apart from thermodynamic equilibrium. At non-zero
frequencies, the expression involves correlation functions of
boundary currents and cannot be put in the standard Green-Kubo form
involving currents inside the system.
\end{abstract}

\maketitle

\section{Introduction}
\label{sec:intro}

One of the important tools in the study of transport phenomena is
the Green-Kubo formula\cite{green,kubo2}, which relates the
equilibrium correlation function of any two conserved currents (i.e.
currents associated with conserved charges) to
the linear response of the first when a gradient in the potential
conjugate to the second is applied.  The response is expressed in
terms of conductivity coefficients of a system, in the thermodynamic
limit when boundary effects can be neglected.  The proof\cite{kubo2}
relies on Onsager's\cite{onsager} relation
between the time evolution of equilibrium and non-equilibrium
fluctuations, or obtains the response to an external field\cite{kubo1}
and relates it to the response to an internal gradient\cite{luttinger}.
One can thence obtain the conductance of a system if it is sufficiently
large and if the conductivity does not diverge.

If either of these conditions breaks down, one has to find the conductance
of the system directly. This can be done by connecting it
to infinite reservoirs\cite{ford,fisher}, but requires
assumptions about the reservoirs and the appropriate ones can be
subtle\cite{szafer}. Alternatively, a Green-Kubo like formula has been
obtained\cite{andrieux} for the conductance(s) of an open classical system
without having to deal with the reservoirs. This is done by assuming that
the system is in a non-equilibrium steady state and (implicitly) that
the current driven by the external potential gradient is a scalar. The
assumptions restrict the result to the zero-frequency conductance of a
quasi one-dimensional (in a sense to be made precise later) system.

In this paper, we derive a formula for the generalized linear
conductance of a finite classical system without any of these
restrictions.  The system can have an arbitrary shape and number
of reservoirs, and the formula applies at any frequency.  The only
assumption made about the reservoirs is that they are in thermodynamic
equilibrium.  The coupling between the system and the reservoirs
is assumed to be such that if the system starts in equilibrium at
the same thermodynamic potentials as all the reservoirs, it remains
in equilibrium, but we do not assume that the reservoirs can equilibrate
the system. For a quasi one-dimensional system at zero frequency,
the formula can be transformed into that of Ref.~\cite{andrieux}
which is of the standard Green-Kubo form, i.e. it involves the
equilibrium fluctuations of currents inside the system. We show
that this transformation is not possible at finite frequencies.

The derivation here builds on previous work that obtained the thermal
conductance at zero\cite{dkn} and non-zero\cite{dnks10} frequency
of a finite classical system in contact with heat baths. Although
a large variety of heat baths were considered, the proof had to be
painstakingly constructed separately for each bath. Thus it was not
clear whether it might fail for some types of heat baths and whether
it applied to other conserved currents. The present paper resolves
these questions.

The rest of this paper is organized as follows. In
Section~\ref{sec:whitenoise} , we briefly review the result of
Ref.~\cite{dkn,dnks10} for the thermal conductance of a Fermi-Pasta-Ulam
(FPU) chain with Langevin heat baths at the ends.  The notation is more
general and part of the proof is slightly different to allow it to be
extended to other conserved currents.  In Section~\ref{sec:particle},
we extend the proof to the case of all conserved currents. In
Section~\ref{sec:dc} we transform the formula at zero frequencies to
the `standard' form, and discuss why this is not possible at non-zero
frequencies.

\section{FPU chains with Langevin baths}
\label{sec:whitenoise}
We first review the derivation of Refs.~\cite{dkn,dnks10} for the 
heat conductance
of a $N$-particle FPU chain with Langevin baths at the ends.
The equations of motion are
\begin{eqnarray}
m_l \dot v_l &=& 
-\frac{\partial}{\partial x_l} [U(x_{l-1} - x_l) + U(x_l - x_{l+1}) + V(x_l)]\nonumber\\
&+& \delta_{l,1}[\eta_L(t) -\gamma_L v_1] + \delta_{l,N}[\eta_R(t) -\gamma_R v_N]
\label{dyneq}
\end{eqnarray}
for $l = 1, 2,\ldots N.$ Here $m_l,x_l,v_l$ are the mass, position and velocity of the 
$l$'th particle. $U$ and $V$ are the interparticle and onsite potentials with $x_0 = x_{N+1} = 0.$ 
$\gamma_{L,R}$ and $\eta_{L,R}(t)$ are the damping and Gaussian noise from 
reservoirs at temperatures $T_{L,R},$ satisfying
\begin{equation}
\langle \eta_{L,R}(t)\eta_{L,R}(t^\prime)\rangle = 
2 \gamma_{L,R} k_B T_{L,R}\delta(t-t^\prime).
\end{equation}

In the first stage of the proof, the Fokker Planck equation for the full
phase space distribution function $P(\bx; \bv; t)$  is constructed, where
$\bx = \{x_1\ldots x_N\}$ and $\bv = \{v_1\ldots v_N\}.$ 
If $T_L = T_R,$,
the steady state solution to the equation is the equilibrium Boltzmann
distribution $P^0(\bx,\bv).$ For $T_{L,R} = T \pm \half\Delta T,$ we have
\begin{equation}
\frac{\partial P}{\partial t} = 
= \hat L P + \hat L^{\Delta T} P
\label{firstorder}
\end{equation}
where $\hat L$ is the equilibrium ($\Delta T = 0$) Fokker Planck operator 
\begin{eqnarray}
\hat L &=& 
\frac{\gamma_L}{m_1}\left[\frac{k_B T}{m_1}\frac{\partial^2}{\partial v_1^2} + \frac{\partial}{\partial v_1} v_1\right] + (1,L)\rightarrow(N,R) \nonumber\\
&+& 
\sum_i \left[-v_i\frac{\partial}{\partial x_i} - F_i \frac{\partial}{\partial v_i}\right] 
\end{eqnarray}
($F_i$ is the force on the $i$'th particle) and 
\begin{equation}
\hat L^{\Delta T} = 
\frac{k_B\Delta T}{2} \bigg[\frac{\gamma_L}{m_1^2}\frac{\partial^2}{\partial v_1^2} 
- \frac{\gamma_R}{m_N^2}\frac{\partial^2}{\partial v_ N^2}\bigg].
\label{LDeltaT}
\end{equation}
With $P(\bx,\bv, t) = P^0 + p(\bx,\bv, t),$ to linear order in
$\Delta T$ 
\begin{equation}
p({\bf x};{\bf v}; t) = 
\int_{-\infty}^t e^{(t - t^\prime)
\hat L}J^e_{fp}({\bf v}) P^0({\bf x}, {\bf v}) \Delta\beta(t^\prime) dt^\prime
\end{equation}
where $\Delta\beta = \Delta(1/k_B T) = -\Delta T/(k_B T^2)$ and $J^e_{fp}({\bf v})$ is defined by 
\begin{equation}
\frac{\partial P}{\partial t}\bigg\vert_{P=P^0} = \hat L^{\Delta T} P^0 =
(\Delta\beta) J^e_{fp} P^0 
\label{jfpdef}
\end{equation}
(the superscript in $J^e_{fp}$ referring to the energy).
For any observable $A,$ we define $\langle\delta A\rangle = \langle A\rangle - \langle A\rangle_0$ in terms of its expectation values with $\Delta\beta\neq 0$ and $\Delta\beta = 0.$ Then
\begin{equation}
\langle \delta A(t_1)\rangle_{\Delta T} = 
\int_{-\infty}^{t_1} \langle A(t_1) J^e_{fp}(t_2)\rangle_{eq} \Delta\beta(t_2) dt_2
\label{expval}
\end{equation}
where the correlation function on the right hand side is evaluated in
equilibrium with $\Delta T = 0.$ 

More generally, if the reservoirs at the ends have different values
for some thermodynamic potential $\Phi^\rho$ whose conjugate conserved
density is $\rho,$ then
\begin{equation}
\langle \delta A(t_1)\rangle_{\Delta \Phi} = 
\int_{-\infty}^{t_1} \langle A(t_1) J^\rho_{fp}(t_2)\rangle_{eq} \Delta\Phi^\rho(t_2) dt_2
\label{expvalg}
\end{equation}
with $J^\rho_{fp}$ defined by the generalization of Eqs.(\ref{firstorder})
and (\ref{jfpdef}).  For example, for the particle current, $\Phi^n =
-\beta\mu,$ where $\mu$ is the chemical potential.

The second part of the proof is specific to Langevin baths at different
temperatures, and is slightly different from Refs.~\cite{dkn,dnks10}. From the
Fokker Planck equation, one can verify that
\begin{equation}
J^e_{fp}  = {\gamma_R\over {2 m_N}} [m_N v_N^2 - k_B T]
- \frac{\gamma_L}{2 m_1}[m_1 v_1^2 - k_B T].
\label{jfpdef1}
\end{equation}
We define the boundary energy current variable $J^e_b$ as the mean of
the instantaneous energy currents flowing into the system from the left
reservoir and flowing out of the system to the right reservoir. Thus
\begin{equation}
J^e_b(t) = \half (j^e_{1,L} - j^e_{N,R})
\label{jbdef}
\end{equation}
where
\begin{eqnarray}
j^e_{1,L}(t) &=& -\gamma_L v_1^2(t) + \eta_L(t) v_1(t),\nonumber\\
j^e_{N,R}(t) &=& -\gamma_R v_N^2(t) + \eta_R(t) v_N(t).
\label{lang}
\end{eqnarray}
Then we will prove the relation
\begin{equation}
\langle A(t_1) J^e_{fp}(t_2)\rangle_{eq} = 
-\langle A(t_1) J^e_b(t_2)\rangle_{eq}\qquad{\rm if}\,\, t_1 > t_2.
\label{jbjfp}
\end{equation}
To prove Eq.(\ref{jbjfp}), we use Eq.(\ref{lang}) and average over the
noise. In performing this average, we discretize the equations of motion
Eq.(\ref{dyneq}) in the usual manner, with the right hand side evaluated
at time $t$ and the left hand side equal to $m_l[v_l(t+\epsilon) -
v_l(t)]/\epsilon.$ The boundary heat currents are
\begin{eqnarray}
j^e_{1,L}(t)&\equiv& \half[v_1(t) + v_1(t+\epsilon)][\eta_L(t) - \gamma_L v_1(t)]
\nonumber\\
j^e_{N,R}(t)&\equiv& \half[v_N(t) + v_N(t+\epsilon)][\eta_R(t) - \gamma_R v_N(t)].
\label{jdisc}
\end{eqnarray}
Usually, the noise average is performed at fixed $v_1(t),$ yielding 
\begin{equation}
\langle j^e_{1,L}(t)\rangle_\eta = \gamma_L (k_B T/m_1 - v_1^2(t)).
\label{jnoiseavg}
\end{equation} 
Naively, this would seem to yield Eq.(\ref{jbjfp}) without the minus sign 
on the right hand side.

However, we want to use the Fokker-Planck evolution operator
over the interval $t_1 > t > t_2$ to evaluate the left hand
side of Eq.(\ref{jbjfp}). Therefore, we use $J^e_b(t_2-\epsilon)$
in Eq.(\ref{jbjfp}) and perform the noise average for fixed
$v_{1,N}(t_2),$ not $v_{1,N}(t_2 - \epsilon).$ We change variables
in Eq.(\ref{jdisc}) from $\eta_{L,R}(t)$ to $\tilde\eta_{L,R}(t)
= \eta_{L,R}(t)-2\gamma_{L,R} v_{1,N},$ since $\eta_{L,R}(t)$ are
uncorrelated with $v_{1,N}(t+\epsilon).$ It is easy to verify that
\begin{eqnarray}
\langle j^e_{1,L}(t-\epsilon) \rangle_\eta &=& \gamma_L(v_1^2(t) - k_B T/m_1),
\nonumber\\
\langle j^e_{N,R}(t-\epsilon)\rangle_\eta &=& \gamma_R(v_N^2(t) - k_B T/m_N).
\label{jnoiseavg1}
\end{eqnarray}
The same result can also be obtained by using the fact that evolving
backwards in time from $t_2$ reverses the currents, and from the
time reversal invariance of the dynamics the noise-averaged time
reversed current is given by Eq.(\ref{jnoiseavg}).  Comparing with
Eq.(\ref{jfpdef}), the noise averaged $J^e_b(t_2-\epsilon)$ is equal
to $-J^e_{fp}(t_2),$ thus proving Eq.(\ref{jbjfp}). Combining Eqs.(\ref{expvalg}) and (\ref{jbjfp}), we have
\begin{equation}
\langle\delta A(t_1)\rangle_{\Delta T} = -\int_{-\infty}^{t_1}\langle A(t_1) J^eb(t_2)\rangle_{eq}\Delta\beta(t_2) dt_2
\label{refeq2}
\end{equation}
which gives the response to a general time dependent (i.e. non-zero
frequency) variation in the temperatures of the reservoirs. Equivalently,
Fourier transforming, 
\begin{equation}
\langle\delta A(\omega)\rangle_{\Delta T} = -\Delta\beta(\omega)
\Bigg[\int_0^\infty\exp[i\omega t] \langle A(t) J_b^e(0)\rangle_{eq} dt\Bigg].
\end{equation}
This equation is 
now generalized to all conserved currents in Section~\ref{sec:particle}.

\section{Generalized currents}
\label{sec:particle}
We first consider the case of particle currents. Although 
Section~\ref{sec:whitenoise} was for a lattice
system with a fixed number of particles, it can be easily
extended to a system with a continuous coordinate $x$ in which the particles
move around, allowing particles to be enter and leave the system from the 
reservoirs.

Because the number of particles in the system is no longer conserved,
we work in the grand canonical ensemble. Let $P_N(\bx, \bv, t)$ be the
probability density for the system to be in an $N$-particle configuration
with coordinates $(\bx, \bv)$ at time $t.$ Thus $\int P_N(\bx,\bv, t)
d\bx d\bv$ is the probability for the system to be in an $N$-particle
configuration.  In equilibrium, $\int P_N^0(\bx, \bv) d\bx d\bv =
\exp[\beta\mu N] Z_N/\Omega,$ where $Z_N$ is the canonical partition
function for the $N$-particle system and $\Omega$ is the grand partition
function. The Fokker Planck probability density is
now an infinite column vector, and the time evolution operators $\hat L$
and $\hat L^{\Delta \mu}$ are matrices. We will denote the column vector
as $P(\bx, \bv, t).$ Analogous to Eq.(\ref{jfpdef}), we define $J_{fp}^n$ 
through
\begin{equation}
\frac{\partial P}{\partial t}\bigg\vert_{P=P^0} = \hat L^{\Delta \mu} P^0 =
(-\Delta\beta\mu) J^n_{fp} P^0.
\label{jfpdef_n}
\end{equation}
The operator $\hat L^{\Delta\mu}$ is due to the difference in the
chemical potentials of the reservoirs and the system, and changes the number of
particles $N.$

Let $\lambda^L_{N^\prime, N}(\beta\mu_L)$ be the transition
rate from $P_N(x_1\ldots v_N)$ to $P_{N^\prime}(x^\prime_1\ldots
v^\prime_{N^\prime})$ due to the left reservoir.  (The dependence of
$\lambda$ on other variables such as $\bx,\bv$ and $t$ 
is implicit; if some of the reservoir variables are so slow that they cannot be integrated out to yield transition rates, one can augment the arguments of $P_N$ and the $\lambda$'s to include these variables.) The contribution from the left reservoir to the
first order equation for $\Delta P_N = P_N - P_N^0$ is
\begin{eqnarray}
\partial_t^L \Delta P_N(\bx_1\ldots \bv_N) 
&=& \Delta(\beta\mu)_L\sum_{N^\prime} 
\int \Big[(\partial_{\beta\mu_L}\lambda^L_{N, N^\prime}) P^0_{N^\prime} \nonumber\\
 &-& (\partial_{\beta\mu_L}\lambda^L_{N^\prime, N}) P^0_N\Big] 
  d\bx^\prime_1\ldots d\bv^\prime_{N^\prime}
\label{rateeq}
\end{eqnarray}
where the derivatives on the right hand side are taken at $(\beta\mu)_L=
\beta\mu.$ Here $\partial_t^L\Delta P_N$ is the part of the time
evolution of $\Delta P_N$ due to the left reservoir; $\partial_t
P_N = (\partial_t^L + \partial_t^R) P_N.$

Since the left reservoir cannot disturb the grand canonical ensemble
distribution when it is at the same $\beta\mu$ as the system, we
have
\begin{equation}
\sum_{N^\prime}\int\lambda^L_{N,N^\prime}P^0_{N^\prime} 
d\bx^\prime_1\ldots d\bv^\prime_{N^\prime}
  = \sum_{N^\prime}\int\lambda^L_{N^\prime,N} P^0_N 
d\bx^\prime_1\ldots d\bv^\prime_{N^\prime}
\label{balance}
\end{equation}
whenever $(\beta\mu)_L= \beta\mu.$ 
The derivatives on the right hand side of Eq.(\ref{rateeq}) can
then be transferred:
\begin{eqnarray}
\partial^L_t \Delta P_N(\bx_1\ldots \bv_N) 
&=& -\Delta(\beta\mu)_L\sum_{N^\prime} \int 
\bigg[\lambda^L_{N,N^\prime} (\partial_{\beta\mu}P^0_{N^\prime} )\nonumber\\
&& -\lambda^L_{N^\prime,N} (\partial_{\beta\mu}P^0_N)\bigg]
d\bx^\prime_1\ldots d\bv^\prime_{N^\prime}.
\end{eqnarray}
Since $P_N^0\sim\exp[N\beta\mu],$ applying Eq.(\ref{balance}) again
\begin{eqnarray}
\partial^L_t \Delta P_N
&=& \Delta(\beta\mu)_L\sum_{N^\prime} (N - N^\prime)
   \int \lambda^L_{N,N^\prime} P^0_{N^\prime}.
\label{rateeq1}
\end{eqnarray}
Now $\lambda^L_{N,N^\prime} P^0_{N^\prime}$ integrated over
$\bx^\prime_1\ldots\bv^\prime_{N^\prime}$ is the rate at which systems
transition from an $N^\prime$ particle state to an $N$-particle state
at the phase space point $\bx_1\ldots \bv_N.$ Therefore the probability
density for the system to be in an $N$-particle state at time $t$ and
have received $N - N^\prime$ particles from the left reservoir between
time $t-\delta t$ and $t$ is $\lambda^L_{N,N^\prime} P^0_{N^\prime}\delta
t.$ For a system in an $N$-particle state at $\bx_1\ldots \bv_N,$
the particle current from the left reservoir immediately {\it before\/}
the time $t$ is then $j_{1,L}^N P_0^N(\bx,\bv) = \sum_{N^\prime}
(N - N^\prime)\lambda^L_{N,N^\prime} P^0_{N^\prime}.$ The stipulation
that the current has to be evaluated just before $t$ is important 
as seen from the discussion around Eqs.(\ref{jnoiseavg}) and
(\ref{jnoiseavg1}).

Comparing with Eq.(\ref{rateeq1}), including the effect of the
reservoir to the right with $\Delta(\beta\mu)_L = -\Delta(\beta\mu)_R =
\half\Delta(\beta\mu)$ and using Eq.(\ref{jfpdef_n}), we obtain
\begin{equation}
J^n_{fp} P^0 = -J_b^n P^0
\end{equation}
where $J_b^n$ is defined in terms of the particle currents from the
reservoirs in the same manner as Eq.(\ref{jbdef}). 

Although energy currents were handled differently in
Section~\ref{sec:whitenoise}, there is no reason why one could not
have proceeded as we have done here: define $P_E(\bx,\bv,t)$ as the
probability density for the system to have an energy $E$ and be at the
phase-space point $(\bx,\bv).$ (The sums over $N^\prime$ would have been
replaced by integrals over $E^\prime.$) Since the number of variables
in $(\bx,\bv)$ is independent of the energy $E$ and different energies
correspond to non-overlapping regions in phase space, this would have
been an unnecessary complication. But from this we see that although the specific example
of particle currents has been used in this Section,
{\it all\/} conserved currents can be treated
in the same manner, and
\begin{equation}
\langle A(t_1) J^\rho_{fp}(t_2)\rangle_{eq} = 
-\langle A(t_1) J^\rho_b(t_2)\rangle_{eq}\qquad{\rm if}\,\, t_1 > t_2
\label{jbjfpg}
\end{equation}
for any conserved current $J^\rho.$ Nor is one limited to two reservoirs: 
since the proof we have constructed for Eq.(\ref{jbjfpg}) deals with the 
reservoirs separately, one can consider an arbitrary number of reservoirs
at potentials $\Phi^\rho_\alpha.$ Combining with Eq.(\ref{expvalg}), 
\begin{equation}
\langle \delta A(t)\rangle_{\Delta \Phi} = \sum_\alpha 
\int_{-\infty}^t \langle A(t) j^\rho_{b,\alpha}(t^\prime)\rangle_{eq} 
\Delta\Phi_\alpha^\rho(t^\prime)\, dt^\prime
\label{expvalg_m}
\end{equation}
If $A$ is the current $j^\sigma_{b,\kappa}$ at the boundary
$\kappa,$ one
obtains the conductance~\cite{footdnks}: 
\begin{equation}
\langle j^\sigma_{b,\kappa}(t)\rangle_{\Delta \Phi} = \sum_{\alpha ,\rho}
\int_{-\infty}^t \langle j^\sigma_{b,\kappa}(t) j^\rho_{b,\alpha}(t^\prime)\rangle_{eq} 
\Delta\Phi_\alpha^\rho(t^\prime)\, dt^\prime.
\label{expvalg_n}
\end{equation}
As with Eq.(\ref{refeq2}) this can be Fourier transformed to yield
\begin{equation}
\langle j^\sigma_{b,\kappa}(\omega)\rangle_{\Delta\Phi} = 
\sum_{\alpha,\rho} \Delta\Phi_\alpha(\omega) \int_0^\infty \exp[i\omega t] \langle j^\sigma_{b,\kappa}(t) j^\rho_{b,\alpha}(0)\rangle_{eq} dt.
\end{equation}
The equilibrium
correlation function in Eq.(\ref{expvalg_m}) must be evaluated with
the {\it same\/} boundary conditions as the left hand side, i.e. with
reservoirs connected to the open system (but, unlike the left hand side,
with the reservoirs at the same thermodynamic
potentials). Different boundary conditions may be inequivalent even in 
the large system limit~\cite{deutsch,dnks10}.

\section{Zero frequency conductance}
\label{sec:dc}
At zero frequencies, for a quasi one dimensional system with two
reservoirs, Eq.(\ref{expvalg}) can be transformed into a more familiar
form. The proof is a generalization of the one in Ref.~\cite{dkn}. If
the reservoirs are at potentials $\Phi^\rho \pm (\Delta\Phi^\rho)/2,$ from
Eqs.(\ref{expvalg}) and (\ref{jbjfpg}) 
\begin{equation}
\langle \delta A(t)\rangle_{\Delta\Phi}  = \sum_\rho (\Delta\Phi^\rho)
\int_{-\infty}^t \langle A(t) J^\rho_b(t^\prime)\rangle_{eq} dt^\prime.
\label{dc1}
\end{equation}

For concreteness, we first consider a one dimensional lattice.
Let $\rho_i$ be the amount of the conserved quantity $\rho$ at the $i$'th 
particle, $j_{i+1,i}^\rho$ be its current 
between the $i$'th and $i+1$'th particles, and $J^\rho(t) = \sum_{i=0}^{N}
j_{i+1, i}^\rho(t).$ If $A= J^\sigma$ is the current flowing through the
entire chain that is associated with a conserved quantity $\sigma,$
Eq.(\ref{dc1}) becomes
\begin{equation}
\langle \delta J^\sigma(t)\rangle_{\Delta\Phi}  = \sum_\rho (\Delta\Phi^\rho)
\int_{-\infty}^t \langle J^\sigma(t) J^\rho_b(t^\prime)\rangle_{eq} dt^\prime.
\label{dc2}
\end{equation}

Define $D^\rho(t) = \sum_{k=1}^N (N + 1 - 2 k)
\rho_k.$ From the continuity equation,
\begin{equation}
dD^\rho/dt = - 2 J^\rho(t) + 2 (N - 1) J^\rho_b(t).
\label{Dlatt}
\end{equation}
Multiply both sides of this equation by $J^\sigma(t_1),$ take an equilibrium
average, and integrate over $-\infty < t < t_1.$ Then $\langle J^\sigma(t)
D^\rho(-\infty)\rangle_{eq} = \langle J^\sigma\rangle_{eq}\langle
D^\rho\rangle_{eq} = 0,$ and if $\rho$ and $\sigma$ have the same symmetry
under time reversal or the system is symmetric under reflection about
its middle, 
\begin{equation}
\langle J^\sigma(t) D^\rho(t)\rangle_{eq} = 0.
\label{dc_cond}
\end{equation}
If either condition is satisfied for each $\rho$ in
Eq.(\ref{dc2}), 
\begin{equation}
\langle \delta J^\sigma(t)\rangle_{\Delta\Phi}  = 
\frac{1}{N-1}\sum_\rho (\Delta\Phi^\rho)
\int_{-\infty}^t \langle J^\sigma(t) J^\rho(t^\prime)\rangle_{eq} dt^\prime.
\label{dc3}
\end{equation}

For a continuum system of length $L,$ $D^\rho$ is
defined as $\int \rho(x) (L - 2 x) dx,$ and Eqs.(\ref{Dlatt}) and
(\ref{dc3}) are obtained with $N-1$ replaced by $L.$ The factor of $N-1$
in Eqs.(\ref{Dlatt}) and (\ref{dc3}) is replaced with $L.$ In higher 
dimensions, we define $D^\rho = -2\int x\rho(\br) d\br$ and use
\begin{equation}
d D^\rho/dt = 2\int x\nabla\cdot j^\rho(\br) d\br = -2 \int j_x^\rho(\br) 
+2 \sum_{\alpha = 1,2} j^\rho_{b,\alpha} x_\alpha
\end{equation}
where $j_x$ is the $x$ component of the current and the sum is over the reservoirs. Then if Eq.(\ref{dc_cond}) is
satisfied, Eq.(\ref{dc3}) is obtained with $J^\rho\rightarrow J^\rho_x$
on the right hand side and $N-1$ replaced by the $x$-separation
of the two reservoirs\cite{footpeter}.

Eq.(\ref{dc_cond}) is always satisfied for quasi one dimensional
homogeneous systems. By quasi one dimensional, we mean that the system
is a tube whose cross-section does not vary along its length,
with reservoirs at its ends. Translational invariance within the tube
ensures that currents in response to potential differences
between the reservoirs are parallel to the orientation of the tube, and so are effectively
scalar.  This is implicitly assumed in Ref.~\cite{andrieux}, which
derives Eq.(\ref{dc3}) for a non-equilibrium steady state. Eq.(\ref{dc_cond})
is also satisfied if $\rho=\sigma.$ 

Since both particle and energy currents change sign under time reversal,
the continuum version of Eq.(\ref{dc3}) is valid for thermoelectric
transport coefficients:
\begin{eqnarray}
\overline j^n &=& G^{nn} \Delta(\beta\mu) - G^{ne} \Delta\beta\nonumber\\
\overline j^e &=& G^{en}\Delta(\beta\mu) - G^{ee}\Delta\beta
\end{eqnarray}
where $\overline j^n, \overline j^e$ are non-equilibrium steady state
currents flowing in response to differences in the chemical potentials
and temperatures of the reservoirs, and
$G^{\sigma\rho} = \int_0^\infty \langle j^\sigma(t) j^\rho(0)\rangle dt.$

Despite the similarity between Eq.(\ref{dc3}) and the Kubo
formula for the conductivity~\cite{kubo2,kubo1}, there are important
differences.  The caveat at the end of Section~\ref{sec:particle}
is still valid: the correlation function has to be calculated with the
correct boundary conditions. Second, Eq.(\ref{dc3}) has {\it only\/}
been obtained in the zero frequency limit. 

One might wonder whether the $\omega\neq 0$ version of Eq.(\ref{dc3}),
with $\Delta\Phi^\rho(t^\prime)$ taken inside the integral, might
be true even if not proved here. A simple example
proves otherwise: if the contacts to the reservoirs are so weak
that the system is effectively isolated, a non-zero $\Delta\Phi$
produces no response and the boundary current
$j^\rho_{b,\alpha}$ in Eq.(\ref{expvalg_m}) is zero. However,
$\langle J^\sigma(t) J^\rho(t^\prime)_{eq}\rangle\neq 0$ because currents
flow inside an isolated system due to spontaneous fluctuations,
and $\int_{-\infty}^t \langle J^\sigma(t)
J^\rho(t^\prime)_{eq}\Delta\Phi^\rho(t^\prime) dt^\prime\neq 0.$ 
By contrast, if the system is driven by an applied field 
instead of a difference in potentials between the
reservoirs, the $\omega\neq 0$ version of Eq.(\ref{dc3}) is always valid.

To summarize, we have obtained a formula for the
generalized finite-frequency response of a finite classical system
connected to an arbitrary number of reservoirs when the thermodynamic
potentials of the reservoirs are varied. The equilibrium correlation function 
in the formula involves currents flowing in from the reservoirs. At zero frequency,
it reduces to a familiar Green-Kubo form, i.e. it involves equilibrium fluctuations of
the currents inside the system but this is not the case at finite frequencies.


\begin{thebibliography}{}
\bibitem{footdnks}
Eq.(\ref{expvalg_m}) only applies when the quantity $A$ is a function
of the phase space variables of the system; if $A = j^\sigma_{b,\beta},$ the 
thermal conductance of a lattice system connected to Langevin baths has an 
extra term~\cite{dnks10}.
\end{thebibliography}
\end{document}